\documentclass[aps, prmaterials, preprint, superscriptaddress, amsmath, amssymb]{revtex4-2}
\usepackage{graphicx}
\usepackage{color}
\usepackage{caption}
\usepackage{amssymb}
\usepackage{amsmath}
\usepackage{lipsum}
\graphicspath{ {./Figures-main-paper/} }
\usepackage[hidelinks=true]{hyperref}
\usepackage[export]{adjustbox}
\usepackage{multirow}
\usepackage{amsmath}
\usepackage{newunicodechar}
\newunicodechar{ₓ}{\textsubscript{x}}
\usepackage{multibib}
\newcites{Supp}{Supplemental Information References}

\begin{document}
\title{Band Alignment Tuning from Charge Transfer in Epitaxial SrIrO$_3$/SrCoO$_3$ Superlattices}
\author{Jibril Ahammad}
\affiliation{Department of Physics, Auburn University, Auburn, AL, USA}
\affiliation{Department of Materials Science and Engineering, University of Delaware, Newark, DE, USA}

\author{Brian B. Opatosky}
\affiliation{Department of Physics, Auburn University, Auburn, AL, USA}
\affiliation{Department of Physics and Astronomy, University of Delaware, Newark, DE, USA}

\author{Tanzila Tasnim}
\affiliation{Department of Physics, Auburn University, Auburn, AL, USA}
\affiliation{Department of Physics and Astronomy, University of Delaware, Newark, DE, USA}

\author{John W. Freeland}
\affiliation{X-ray Science Division, Argonne National Laboratory, Lemont, IL, USA}

\author{Gabriel Calderon Ortiz}
\affiliation{Department of Materials Science and Engineering, The Ohio State University, Columbus, OH, USA}

\author{Jinwoo Hwang}
\affiliation{Department of Materials Science and Engineering, The Ohio State University, Columbus, OH, USA}
\author{Gaurab Rimal}
\affiliation{Department of Physics, Auburn University, Auburn, AL, USA}
\affiliation{Department of Physics, Western Michigan University, Kalamazoo, MI, USA}

\author{Boris Kiefer}
\affiliation{Department of Physics, New Mexico State University, Las Cruces, NM, USA}

\author{Ryan B. Comes}
\email{comes@udel.edu}
\affiliation{Department of Physics, Auburn University, Auburn, AL, USA}
\affiliation{Department of Materials Science and Engineering, University of Delaware, Newark, DE, USA}


\begin{abstract}
Understanding charge transfer at oxide interfaces is crucial for designing materials with emergent electronic and magnetic properties, especially in systems where strong electron correlations and spin–orbit coupling coexist. SrIrO$_3$/SrCoO$_3$ (SIO/SCO) superlattices offer a unique platform to explore these effects due to their contrasting electronic structures and magnetic behaviors. Building on past theory based on continuity of O 2p band alignment, we employ density functional theory (DFT) to model electron transfer from Ir to Co across the SIO/SCO interface. To characterize these effects, we synthesized epitaxial SIO/SCO superlattices via molecular beam epitaxy. Structural and transport measurements confirmed high crystallinity, metallic behavior, and suppression of Kondo scattering that has been reported in uniform SIO films. Further characterization via X-ray absorption spectroscopy (XAS) revealed orbital anisotropy and valence changes consistent with interfacial charge transfer. Co $K$- and $L_{2,3}$-edge and Ir $L_2$-edge spectra verified electron donation from Ir to Co, stabilizing the perovskite SCO phase and tuning the electronic structure of SIO via hole-doping. O $K$-edge XAS showed band alignment shifts in the SIO layer consistent with DFT predictions. Our work here provides a pathway for engineering oxide heterostructures with tailored magnetic and electronic properties.    
\end{abstract}

\maketitle

\section{Introduction}

Superlattices are a class of engineered thin film heterostructures composed of alternating layers of two or more dissimilar materials. These periodic structures exploit quantum confinement and interfacial effects, leading to emergent physical properties not present in the bulk constituents \cite{gu2024superorders,tsu2007applying,bottner2006aspects}. In a complex oxide superlattice, charge transfer may occur across the interface in cases where cations have different electronegativities. Consequently, the properties of the interface and nearby atomic layers can be fundamentally different from the constituent layers due to a modification of the d-orbital occupancy\cite{chen2017charge,zhong2017band}. 

Interfacial charge transfer has been experimentally established as a powerful mechanism to induce emergent interfacial ferromagnetism in superlattices composed of antiferromagnetic and paramagnetic layers \cite{takahashi2001interface,bhattacharya2008metal,nichols2016emerging, hoffman2016oscillatory}, and high-temperature superconductivity \cite{gozar2008high}. Interfacial 3d/5d systems such as SrCoO$_{3}$ (SCO) and SrIrO$_{3}$ (SIO) offer an excellent platform to explore the rich physics arising from interplay of strong electron correlation and ferromagnetism in cobaltates \cite{takami2014functional} and strong spin–orbit coupling (SOC) in iridates systems \cite{nie2015interplay, zeb2012interplay}. Charge transfer has been observed in other SIO heterostructures, including those with SrMnO$_{3}$ (a 3d antiferromagnetic insulator) \cite{nichols2016emerging,bhowal2019electronic} and SrRuO$_{3}$ (a 4d correlated Weyl ferromagnet) \cite{nelson2022interfacial}. Dimensionality confinement in SrTiO$_3$/SIO superlattices has also been explored to probe the tunability of magnetism in SIO in quasi-2D layers \cite{Hao2017,Kim2016}. However, studies of 3d-5d superlattices and heterostructures where both constituent layers are nonpolar metallic oxides remain largely unexplored. Investigation of the SIO/SCO system can serve as an important step toward bridging this gap.

SCO offers several advantages over past materials grown at the interface with SIO, as it is a ferromagnetic (FM) metal with a Curie temperature between 280 K - 305 K \cite{bezdicka1993preparation, balamurugan2007charge, long2011synthesis}. It crystallizes in a cubic perovskite structure  in the space group $Pm\bar3m$.\cite{long2011synthesis}. Although SrCoO$_{3}$ thin films are widely synthesized to study its intrinsic properties, its metastable nature poses significant challenges in its potential for real-life applications. Due to the multivalent nature of Co, the perovskite (P-SCO) phase can readily degrade into the Brownmillerite SrCoO$_{2.5}$ (BM-SCO) phase, which is an antiferromagnetic (AFM) insulator \cite{schoffmann2021stoichiometric, zhang2019origin}. Some studies report AFM–FM transitions at 2\% \cite{petrie2016strain} and 3\% \cite{callori2015strain} tensile strain, whereas others determined a ferromagnetic ground state even under 3\% tensile strain \cite{wang2020robust}. These discrepancies indicate that SrCoO$_{3}$ lies near a strain-tunable magnetic phase boundary and suggest that alternative mechanisms like charge transfer have merit to tune the magnetic phases.

SIO is a paramagnetic semi-metal  crystallizing with $Pnma$ structure, representing the end member of the Sr$_{n+1}$Ir$_n$O$_{3n+1}$ Ruddlesden-Popper series with n=$\infty$ \cite{cao2018challenge}. A unique feature of iridates is the presence of strong SOC, which competes with crystal field, electron correlations, and other interactions leading to highly tunable and novel electronic and magnetic ground states\cite{cao2018challenge, kim2008novel}. The delicate balance of these interactions is influenced by local atomic configuration \cite{moon2008dimensionality}, structural dimensionality \cite{kim2012dimensionality}, and chemical doping \cite{de2015collapse, wang2011twisted}. Studying charge transfer in SIO superlattices therefore provides a mechanism to probe the interplay of the competing interactions. As an analogy, double perovskite Sr$_2$CoIrO$_{6}$ (SCIO) can be viewed as a superlattice composed of alternating SrCoO$_{3}$ and SrIrO$_{3}$ layers in the (111) direction \cite{wu2021electronic}. A noticeable charge transfer from Ir to Co in the SCIO double perovskite has been reported \cite{wu2021electronic}, resulting in mixed valence state comprising predominantly Ir$^{5+}$ ($J_{eff}$ = 0) and Co$^{3+}$ (S = 2) with a small portion of ($\sim$10\%) Ir$^{6+}$ (S = \(\frac{3}{2}\)) and Co$^{2+}$ (S = \(\frac{3}{2}\)) \cite{esser2018strain,agrestini2019nature}.

There are established frameworks such as Anderson's or the Schottky-Mott rule \cite{zhong2017band,monch2024schottky} to explain charge transfer in semiconductors. However, the application of these rules to transition metal oxides remains challenging because work functions in TMOs are extremely sensitive to specific surface terminations and microscopic details of the surface\cite{zhong2016tuning}. Strong correlations in TMOs make calculating electronic structure more challenging, leading to a search for a more general theory \cite{chen2017charge}. Zhong and Hansmann pursued this goal by setting continuous oxygen matrix in the interface of two perovskites ABO$_3$/AB’O$_3$ and developed a model to predict both magnitude and direction of charge transfer in the interface \cite{zhong2017band}. The authors' model suggests  that in a superlattice consisting of SCO and SIO, Ir will donate electrons to Co, as in past experiments on Sr$_2$CoIrO$_{6}$ \cite{wu2021electronic}. Motivated by their work, we investigated SrIrO$_3$/SrCoO$_3$ superlattices computationally using density functional theory (DFT) modeling and experimentally through molecular beam epitaxy (MBE) synthesis and spectroscopic characterization. Our work confirmed charge transfer phenomena using hard X-ray absorption spectroscopy (XAS) of Co $K$- and Ir $L$-edges on both systems. Our experimental findings are an important step in understanding charge transfer in interface of TMOs and help design novel systems. 
    
\section{Computational Modeling}

To determine the expected degree of charge transfer in these systems, we employed DFT calculations using the Vienna \textit{ab-initio} Simulation Package \cite{kresse1996efficiency, kresse1996efficient} accounting for electronic exchange and correlations within the R2SCAN meta-functional \cite{furness2020accurate}. This meta-functional has been shown to improve thermochemistry \cite{kothakonda2022testing} and electronic structure \cite{swathilakshmi2023performance}. Electrons are treated within the projector augmented-wave framework (PAW) \cite{blochl1994projector, kresse1999ultrasoft}. We followed Material Project recommendations, and adopted a planewave energy cutoff of E\textsubscript{cut} = 520 eV, \cite{jain2013commentary} and a $\Gamma$-centered k-point grid spacing of 0.25 $Å^{-1}$. We optimize SCO and SIO perovskite in the experimentally known cubic ($Pm\bar3m$) \cite{long2011synthesis} and orthorhombic ($Pnma$) \cite{cao2018challenge} equilibrium crystal structures. In preparation for the heterostructure simulations, we transformed the cubic structure in the $Pnma$ structure by a 45$\textdegree$ rotation in the basal plane and cell-doubling along the [010] direction and found that the energies between the two phases is $\sim$1.0 eV/atom, suggesting that the computations are converged. We find that SCO is a ferromagnetic metal (Table \ref{tab:DFT-parameters}). We also tested the effect of octahedral rotation and place SCO in our computed SIO equilibrium structure, and not unexpectedly found that the cubic phase is significantly more stable, and octahedral rotation increases the Co magnetic moment by almost 25\%. This rotation-induced increase in magnetism is expected: in cubic SCO Co 3d and O 2p orbitals shows the highest overlap. With increasing octahedral rotation, this overlap diminishes, leading to more localized Co 3d orbitals, increasing the magnetic moment as computed (Table 1). SIO is also metallic and we find a sizable magnetic moment of $\sim 1 \mu_B$/Ir, consistent with a Ir$^{4+}$ in low-spin configuration. However, after including SOC, the magnetic moment reduces to $\sim 0.1 \mu_B$, in better agreement with the absence of permanent magnetism in SIO\cite{cao2018challenge}. In contrast, we find that SOC has little effect on the magnetism in SCO (Table \ref{tab:DFT-parameters}).

\begin{table}[htbp]
    \centering
\caption{R2SCAN crystallographic parameters of SCO and SIO}
\label{tab:DFT-parameters}
    \begin{tabular}{|l|c|c|c|}
        \hline
        & \textbf{SCO} & \textbf{SCO} & \textbf{SIO} \\ \hline
        Phase& $Pm\bar3m$ & $Pnma$ & $Pnma$ \\ \hline
        Lattice& 3.807 & 5.608 & 5.608 \\ 
        & (3.8289,\cite{long2011synthesis}) & 7.902 & 7.902 \\ 
        & (3.89,\cite{jain2013commentary}) & 5.617 & 5. \\ \hline
        Bandgap& metallic & metallic & metallic \\ \hline
        Rotation angle (deg) & 180.0 & 154.7 & 154.7 \\ \hline
        Magnetic moment ($\mu$$_B$)& 2.856 & 3.781 & 0.990 \\ \hline
        SOC magnetic moment ($\mu$$_B$)& 2.910 & 0.150 & 0.116 \\ \hline
        Formation energy (eV/atom)& 0.000 & 0.158 & NA \\ \hline
    \end{tabular}
\end{table}

The primary interest of this study is to understand the charge transfer in SIO/SCO heterostructures. Heterostructures were generated by stacking $Pnma$ unit cells along [010] direction and replacing  IrO$_2$ layers with CoO$_2$ layers. In the computations, the in-plane lattice parameter was fixed at the value of (La,Sr)(Al,Ta)O$_3$, a = 3.868 Å \cite{chakoumakos1998thermal}, the substrate used in experiment, and the z-lattice parameter was allowed to relax. We distinguish two types of oxygen ions, four equatorial (in-plane, IP) O ions in the TM-O$_6$ octahedra and two bridging oxygen ions to the Sr-O layers (out-of-plane, OOP). All results were obtained from SOC corrected computations. To characterize charge transfer we use the site-projected electronic densities of states (DOS). We computed the layer-resolved O 2p shift in the 40 atom simulation cells with eight formula units, following Zhong and Hansmann \cite{zhong2017band}, by integrating the O 2p DOS for each layer to determine the band-center (center of mass). We note that in SCO/SIO heterostructures O 2p orbitals cross the Fermi level, while in Ref. \cite{zhong2017band} the O 2p band is located well below the Fermi level. This integration provides a qualitative measure for individual phases, but comparing our heterostructures with compositional end-members, we can identify changes due to heterostructure formation, identify changes in the TM-O bond network, and derive insights into the charge transfer mechanism, that can be verified and quantified through XAS experiments.


Ir in SIO strained to the LSAT in-plane lattice paramater shows a small residual magnetic moment of 0.14 $\mu$$_B$/Ir and for IP and OOP oxygen, 0.038 $\mu$$_B$ and 0.007 $\mu$$_B$, respectively, and the groundstate remains metallic . For the corresponding SCO endmember, we find that that it is a ferromagnetic metal, where Co carries a magnetic moment of 2.4 $\mu$$_B$/Co, and both IP and OOP oxygen carries a small but non-zero magnetic moment of $\sim$ 0.2 $\mu$$_B$, consistent with Co d - O 2p hybridization. Therefore, the magnetism in both phases suggests significant hybridization, a finding that is corroborated by the site-projected  density of states that shows O 2p states crossing the Fermi level with O 2p holes in both (strained) end-members. For the TMs and oxygen ions we computed layer resolved d-orbital occupancies as well as oxygen hole occupancies (see Supplemental Table \ref{tab:DFT-orbital-40atm}).

\begin{figure}[htbp]
\centering
\includegraphics[width=1\textwidth]{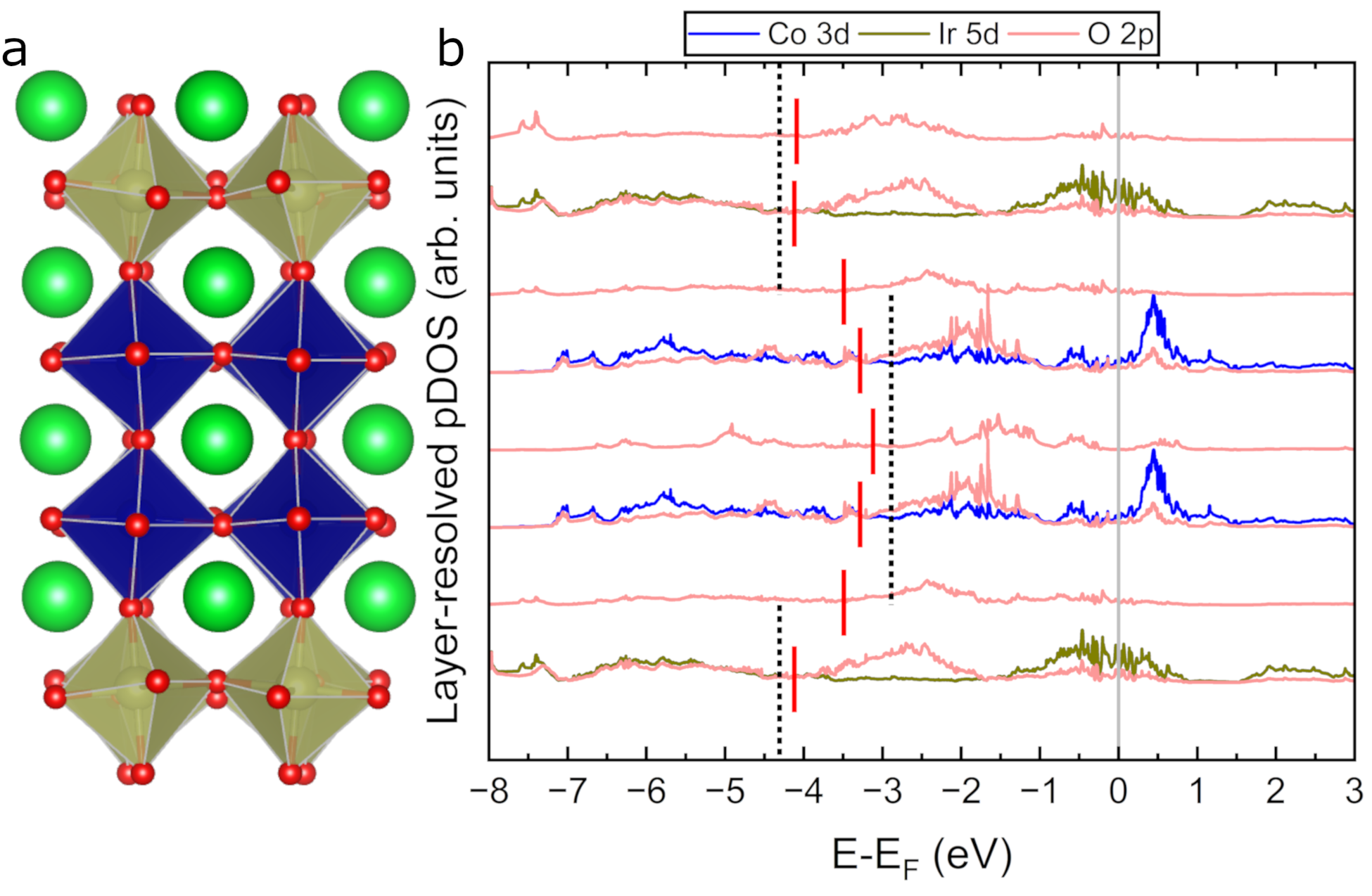}
\caption{(a) Structural model of 2 unit cell (u.c.) SrIrO$_3$/2 u.c. SrCoO$_3$ superlattice with Ir octahedra shown in gold and Co octahedra shown in blue; (b)Layer-resolved partial density of states of O 2p, Co 3d, and Ir 5d. The Fermi level is denoted by gray solid line, superlattice O 2p band centers for each layer are denoted by red lines, and bulk O 2p band centers for SrIrO$_3$ and SrCoO$_3$ are denoted by black dotted lines in the relevant regions.}
\label{PDOS}
\end{figure}

For discussion purposes, we devote particular attention to the 4 Ir/4 Co superstructure, which is equivalent to a 2 unit cell SrIrO$_3$ /2 unit cell SrCoO$_3$ superlattice. The relaxed structure and layer-resolved projected density of states (PDOS) are shown in Figure \ref{PDOS}. By tracking the O 2p band-center relative to the values from our models of strained SIO and SCO\cite{zhong2017band}, we can evaluate expected charge transfer in the superlattices. We see that O 2p in SIO layers within the superlattice is between 0.1 and 0.2 eV closer to the Fermi level than in the bulk system, with no gradient inside the layers. This suggests that SIO is hole-doped due to the interface with SCO but remains metallic and fully screens any electric field. Meanwhile, SCO has a potential gradient as shown by the O 2p band center shifts within the layer, indicating an internal electric field. Collectively, this indicates electron transfer from Ir to Co and would correspond to a more positive oxidation state for Ir (Ir$^{4+x}$) and lower for Co (Co$^{4-x}$). Structural models indicate out-of-plane distortions in both the Ir and Co octahedra, with inequivalent bond lengths along the growth direction that are also consistent with interfacial charge transfer.

\section{Experimental Methods}

Three SIO/[SCO/SIO]$_r$ superlattice films were synthesized using molecular beam epitaxy (MBE) on (001)-oriented (LaAlO$_3$)$_{0.3}$ (Sr$_2$AlTaO$_6$)$_{0.7}$ (LSAT) substrates- a schematic diagram of the structure is shown in Figure \ref{sup-RHEED-SL1-SL2-and-XRD-SL3}a. The films differ in the number of unit cells (uc) in the SIO and SCO layers within individual bilayers as: 1) \textit{SIO/SCO= (4/2) uc} (r= 4 bilayers, 11 nm), 2) \textit{SIO/SCO= (5/3) uc} (r= 4 bilayers, 14.5 nm), 3) \textit{SIO/SCO= (6/4) uc} (r= 3 bilayers, 14 nm). In this work, the films will be referred to as (4/2), (5/3), and (6/4) for notational purposes.  Thus, we studied charge transfer phenomena tuned by a decreasing ratio of SIO: SCO (2:1, 5:3, 3:2 respectively), while total thicknesses remained roughly constant . The top-most SIO layer served as a capping layer to prevent degradation in the underlying SCO layer. The LSAT substrate (a = 3.868 Å) has a lattice mismatch of +1\% with SrCoO$_{3}$ (a$_c$ =3.829 Å) \cite{long2011synthesis, yang2015pressure} and -2.4\% with SrIrO$_{3}$ (bulk pseudocubic lattice parameter $\sim$3.96 Å) \cite{longo1971structure, biswas2017growth, suresh2024tunable}. We used a quartz crystal microbalance (QCM) to calibrate Sr and Co fluxes before loading substrate into the chamber.  We used metallic precursors to generate Sr and Co vapors, and a metal-organic precursor\cite{choudhary2022semi} to supply Ir flux, all of which were evaporated from effusion cells. Before each growth, the substrate was annealed in oxygen plasma at 850 °C for 30 minutes and then cooled to growth temperature (625 °C). SCO layers were grown using a co-deposition method with an oxygen flow rate of 1.5 sccm (P= $1 \times 10^{-5}$ Torr). SIO layers were grown via adsorption-controlled growth with an oxygen flow rate of 0.4 sccm ($6 \times 10^{-6}$ Torr) following our previous recipe \cite{rimal2024strain}. Oxygen plasma from a radio-frequency source was used during the entire process. After growth, the stage was cooled at 20 °C/min. Reference Sr$_2$CoIrO$_{6}$ (SCIO) films were grown on SrTiO$_{3}$ (STO) (100) substrate at 625 °C in an oxygen flow rate of 1.0 sccm  (P= $6 \times 10^{-5}$ Torr), with Sr, Co, and Ir were supplied simultaneously. One of the SCIO samples was further annealed in air at 750 °C for 3 hours in a tube furnace . Additional details on SCIO growth procedures are provided in the supporting information. During each growth, film quality was monitored using in-situ reflection high-energy electron diffraction (RHEED). Based on past experience with Co-based oxides, the beam was blanked during growth of SCO layers to prevent reduction of the material except for 1-2 second intervals each minute to record a picture. 

The crystal structure of the films was characterized using X-ray diffraction (XRD) with a Cu K-$\alpha$ source in the parallel-beam geometry. Film thicknesses and roughnesses were analyzed by X-ray reflectivity (XRR) using the GenX software package \cite{glavic2022genx}.  We performed electrical transport property measurements in a Quantum Design Physical Property Measurement System (PPMS) in the Van der Pauw configuration. High-angle annular dark field (HAADF- STEM) combined with energy-dispersive X-ray spectroscopy (EDX) mapping  was performed in the Center for Electron Microscopy and Analysis at The Ohio State University using an FEI Titan 60-300. 

Polarization-dependent hard X-ray absorption spectroscopy (XAS) on Co $K$ and Ir $L_{2/3}$-edges were performed in the Beamline for Materials Measurement (BMM, 6-BM), National Synchrotron Light Source II (NSLS-II) at Brookhaven National Laboratory (BNL). Measurements were conducted  in fluorescence mode in a 4-element vortex silicon-drift detector using a Si (111) monochromator with step size of 0.5 eV. All measurements were carried out at room temperature. Due to interference from a tantalum transition peak from the LSAT the substrate near Ir $L_3$ energy, we instead measured the Ir $L_2$-edge for the SIO/SCO superlattices and Ir $L_3$-edge for SCIO. Although the Ir $L_3$-edge is conventionally used to determine the Ir oxidation state, Ir $L_2$ white line is still indicative of different oxidation states in Ir compounds \cite{laguna2015electronic, sweers2024epitaxial, duell2021iridium}. As reference standards, we measured Co $K$-edges of LaCoO$_3$ (Co$^{3+}$) films, Ir $L_2$ (and Ir $L_3$) edges of a pure SrIrO$_{3}$ (Ir$^{4+}$) grown on LSAT (SrTiO$_3$). A CoO (Co$^{2+}$) published powder reference was also used. All reference perovskite samples were synthesized in the same MBE chamber as the superlattice/double perovskite samples. Calibration of the measured spectra was performed using metallic Co and gold foils for the Co $K$ and Ir $L$-edges, respectively. XAS data reduction and analysis were performed using the Athena software from the Demeter (version: 0.9.26) library \cite{ravel2005athena}.

In addition, we performed Co L-edge and O $K$-edge XAS measurements of the superlattice samples along with pure SrIrO$_3$ on three different substrates at Beamline 29-ID of the Advanced Photon Source (APS), Argonne National Laboratory. The experimental setup was designed for data collection in both surface-sensitive total electron yield (TEY) and bulk-sensitive fluorescence yield (TFY) modes. In the TEY mode, however, we obtained meaningful data only for one superlattice sample (4/2 uc); for the other two samples, the capping layers were too thick to probe the top cobaltate layer. For the Co L-edge data, we subtracted the background using a combination of a linear pre-edge and an arc-tangent function, followed by normalization with the maximum intensity of each spectrum.

\section{Results and Discussions}

\begin{figure}[htbp]
\centering
\includegraphics[width=1\textwidth]{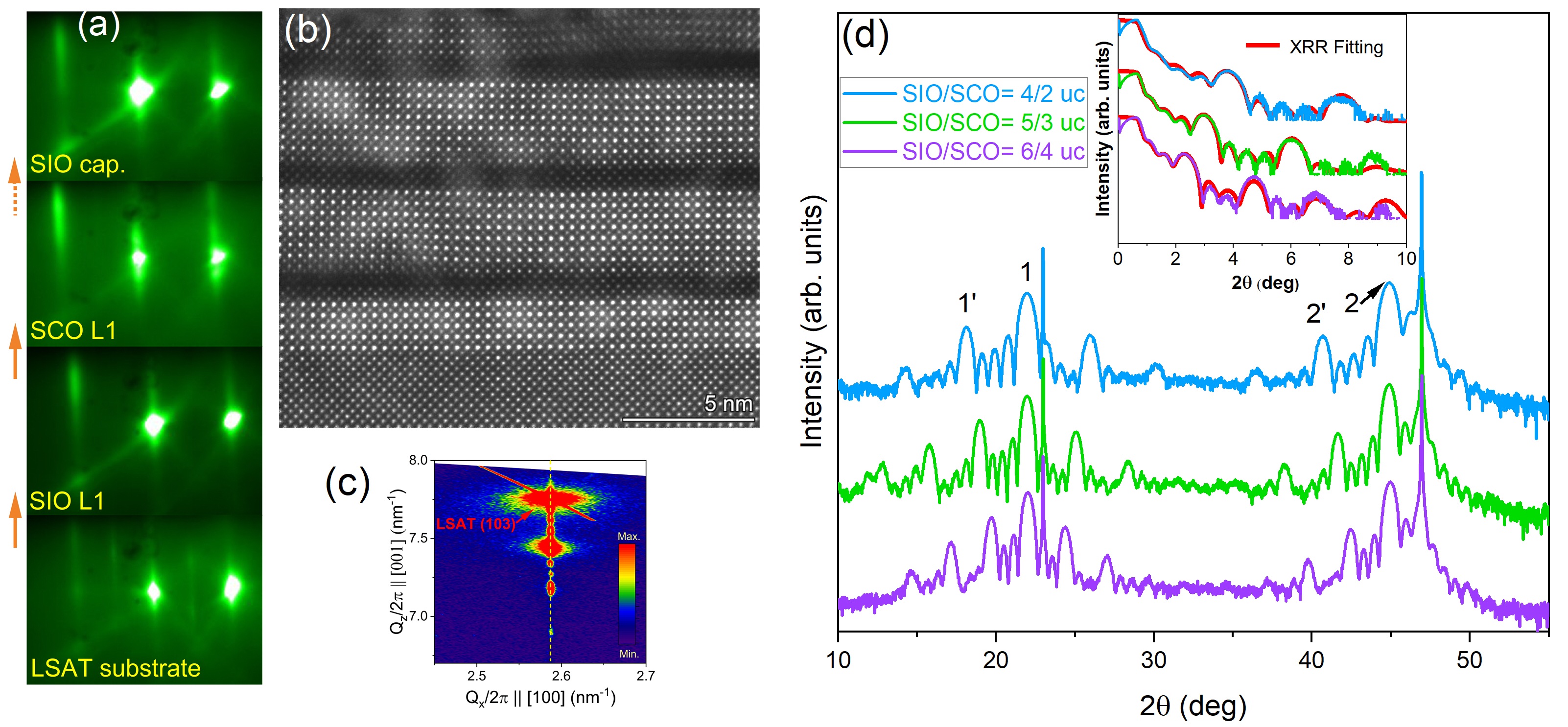}
\caption{(a) RHEED, (b) STEM, (c) RSM of a representative superlattice (SIO/SCO=6/4 uc), (d) XRD of all three superlattices (XRR in inset).}
\label{structure-SL3}
\end{figure}

The crystal structure data for the \textit{SIO, SCO= (6/4) uc} film is presented in Figure \ref{structure-SL3} as a representative for the superlattices. In-situ RHEED (Figure \ref{structure-SL3}(a)) clearly shows the Kikuchi lines and sharp diffraction spots confirming high crystalline quality and atomically flat surfaces. A reciprocal space mapping measurement of this film (Figure \ref{structure-SL3}(d)), performed along the LSAT (103) shows uniform Q$_z$ (marked by the dotted vertical line). This indicates that each layer is locked to the LSAT substrate, confirming the coherently strained and epitaxial nature of the superlattice structure. 

The XRD Bragg peaks (denoted by 1, 2) in Figure \ref{structure-SL3}(c) also demonstrates high-order crystallinity of the films. Superlattice reflections, represented by the satellite peaks (1', 2'), further confirm long-range structural order of the superlattices. XRR simulation suggests an interfacial roughness between 0.1-0.2 nm and a surface roughness between 0.3-0.4 nm. Fits to the superlattice periodicity confirmed that the SCO and SIO layer thicknesses were within half of a unit cell of the nominal values used throughout the paper. Although pure SrCoO$_3$ undergoes significant degradation to the Brownmillerite phase when exposed to air for over 24 hours, XRD data in Fig. \ref{sup-RHEED-SL1-SL2-and-XRD-SL3}(d) indicates that, when incorporated into a superlattice with SrIrO$_3$, it remains stable after 6 months. The structural quality of the superlattices and the perovskite phase of SCO is confirmed by STEM measurements shown in Figure \ref{structure-SL3}(b).

\begin{figure}
  \centering
  \includegraphics[width=0.38\textwidth]{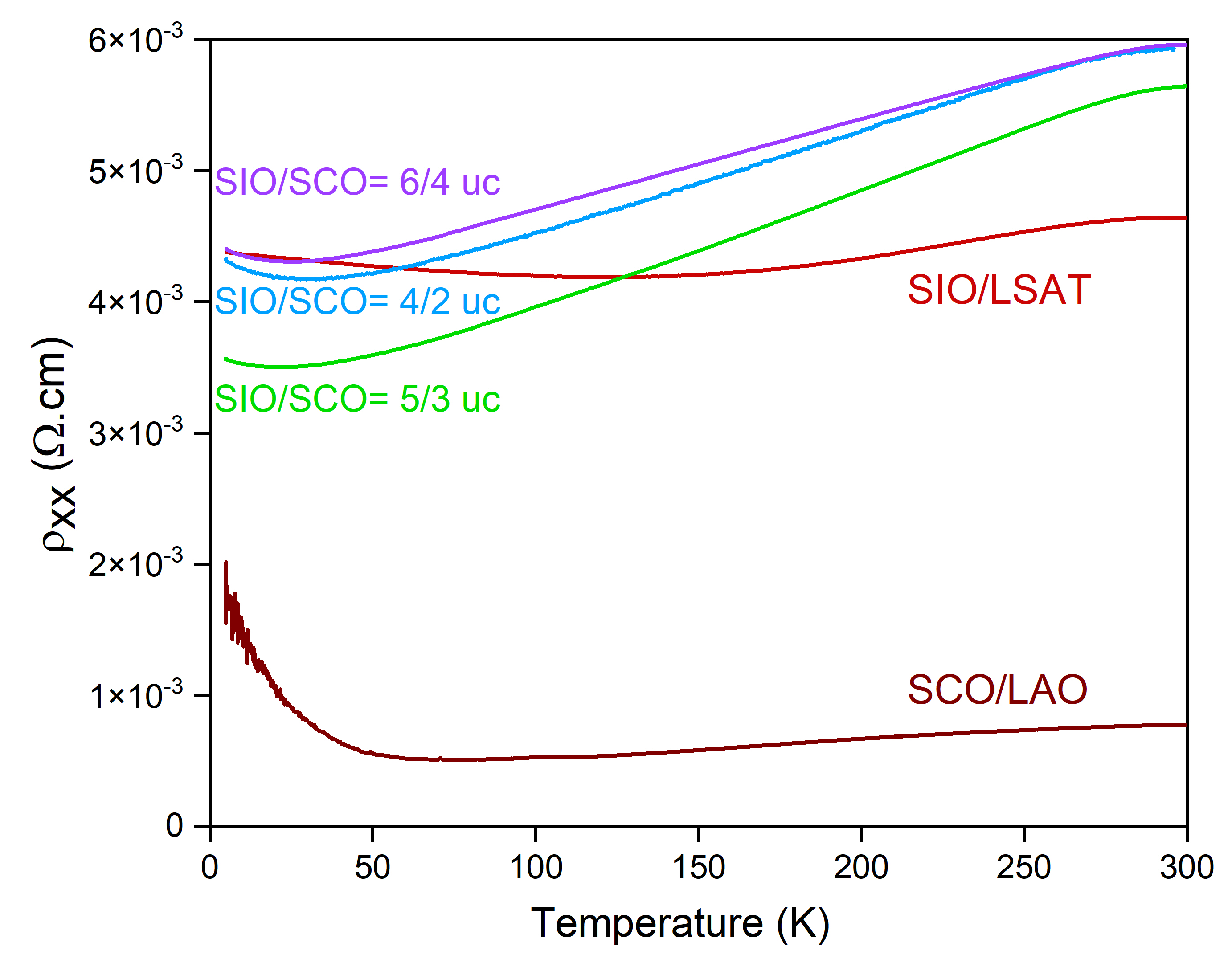}
  \caption{Temperature-dependent resistivity for superlattice samples and reference SrCoO$_3$/LaAlO$_3$ and SrIrO$_3$/LSAT samples.}
  \label{Transport}
\end{figure}

All superlattice films showed metallic behavior, with an upturn below 50 K, as shown in Figure \ref{Transport}. This is likely due to weak localization effects, as no appreciable variation in carrier concentration $\sim 9 \times 10^{21} \, \text{cm}^{-3}$ was found. Notably, while the Kondo effect has been observed in pure SIO grown on various substrates, including LSAT\cite{rimal2024strain}, fits to the resistivity data using localization and scattering models show no contribution of Kondo scattering to the resistivity at low temperature, but do indicate weak localization. In past SIO samples grown via this technique\cite{rimal2024strain,choudhary2022semi}, conduction is dominated by electrons in pure SrIrO$_3$ with strain-induced oxygen vacancies driving Kondo scattering due to dilute concentrations of Ir$^{3+}$ formal charge states. Meanwhile DFT models of the superlattice indicate that SIO layers are expected to donate electrons to SCO, leading to hole doping towards Ir$^{5+}$.  Reference SCO films also showed metallic behavior but with a greater upturn at low temperatures. The more metallic behavior of all three superlattice samples compared to uniform SIO underscores the role of charge transfer in suppressing Kondo scattering. This suggests that the superlattice possesses an emergent electronic state that is distinct from both bulk SIO and SCO.   

\begin{figure}[h]
\centering
\includegraphics[width=1\textwidth]{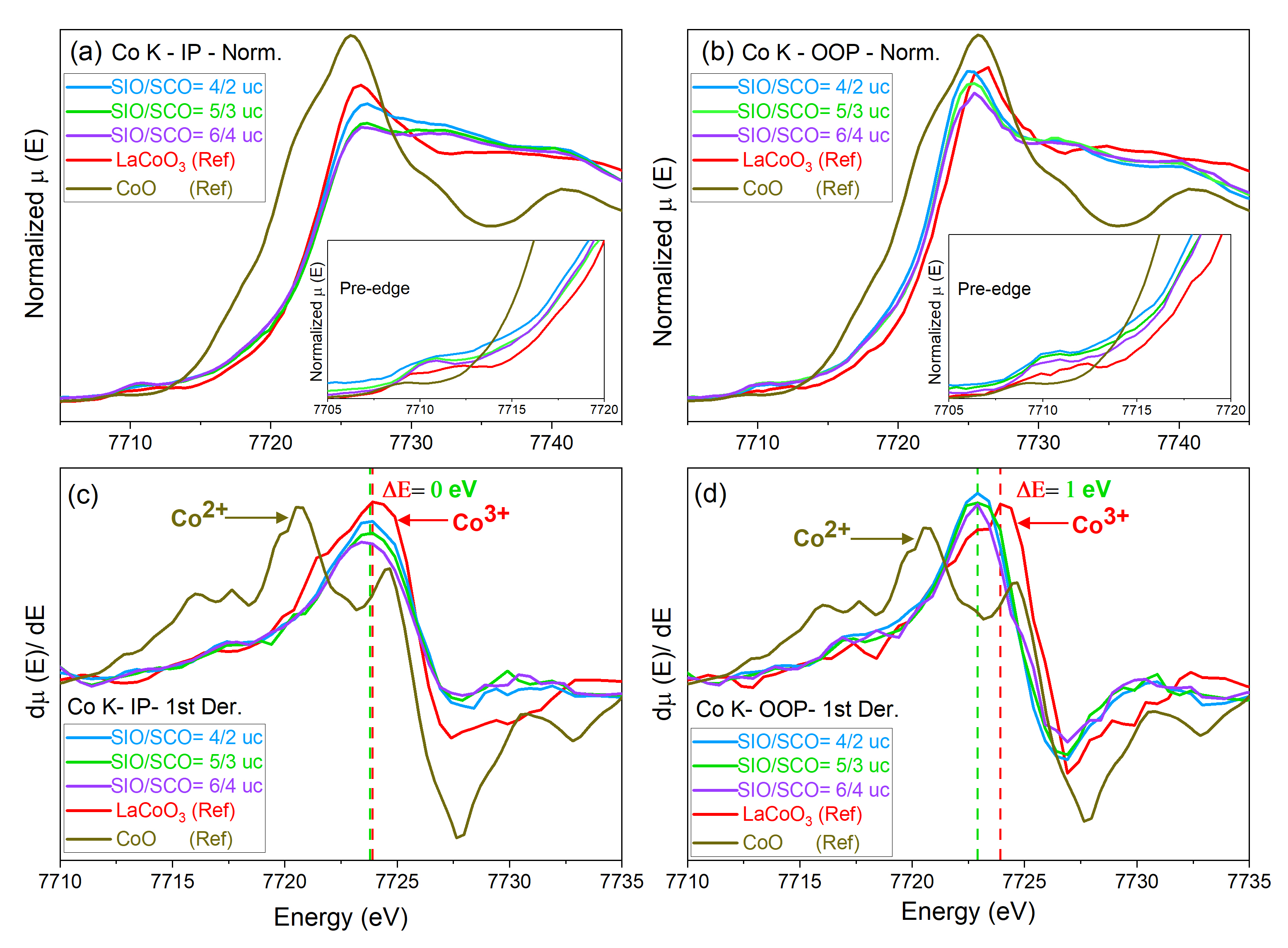}
\caption{XAS Co $K$-edge data for the superlattice films (a), (c) In-plane data (normalized spectra and their first derivatives); (b),(d) Out-of-plane data (normalized spectra and their first derivatives).}
\label{XAS-Co-K-SL}
\end{figure}

To characterize the degree of interfacial charge transfer between Co and Ir, we employed XAS measurements about the Co $K$ and L edges, the Ir $L$-edge, and the O K edge. We begin with discussion of the Co valence states and via the K edge data. To determine the valence, we examine the inflection point of the white line, taken as the maximum of the first derivative of the absorption data  \cite{henderson2014x,deGroot2004}. From the in-plane (IP) polarized XAS data in Figure \ref{XAS-Co-K-SL}(c), the maximum of the first derivative of all superlattices align exactly with LaCoO$_3$, indicating an average of Co$^{3+}$ state. In the out-of-plane (OOP) polarized XAS data, the maximum value of the first derivative for all superlattices shifts further towards lower energy, which would nominally indicate an average Co oxidation state of less than 3+ when considered independently of the IP data. This ambiguity suggests that orbital anisotropy arises due to charge transfer and symmetry breaking at the interface. Given that reference SCO films grown in our system are metallic and exhibit the perovskite phase, indicating oxidation beyond a 3+ formal charge, we argue that Ir from SrIrO$_3$ donates electrons to Co in SrCoO$_3$ which eventually leads to formation of the stable structure for over 6 months, as shown in Supplemental Figure \ref{sup-RHEED-SL1-SL2-and-XRD-SL3}(d). A similar conclusion can be drawn from the Co $K$-edge data of both SCIO samples shown in Supplemental Figure \ref{sup-XAS_SCIO} We observe a greater shift towards lower energy for both IP and OOP data, which most likely originated due to a isotropic charge transfer between Co and Ir atoms in SCIO compared to interactions only at the interfaces of the superlattices.

\begin{figure}
\centering
\includegraphics[width=0.5\textwidth]{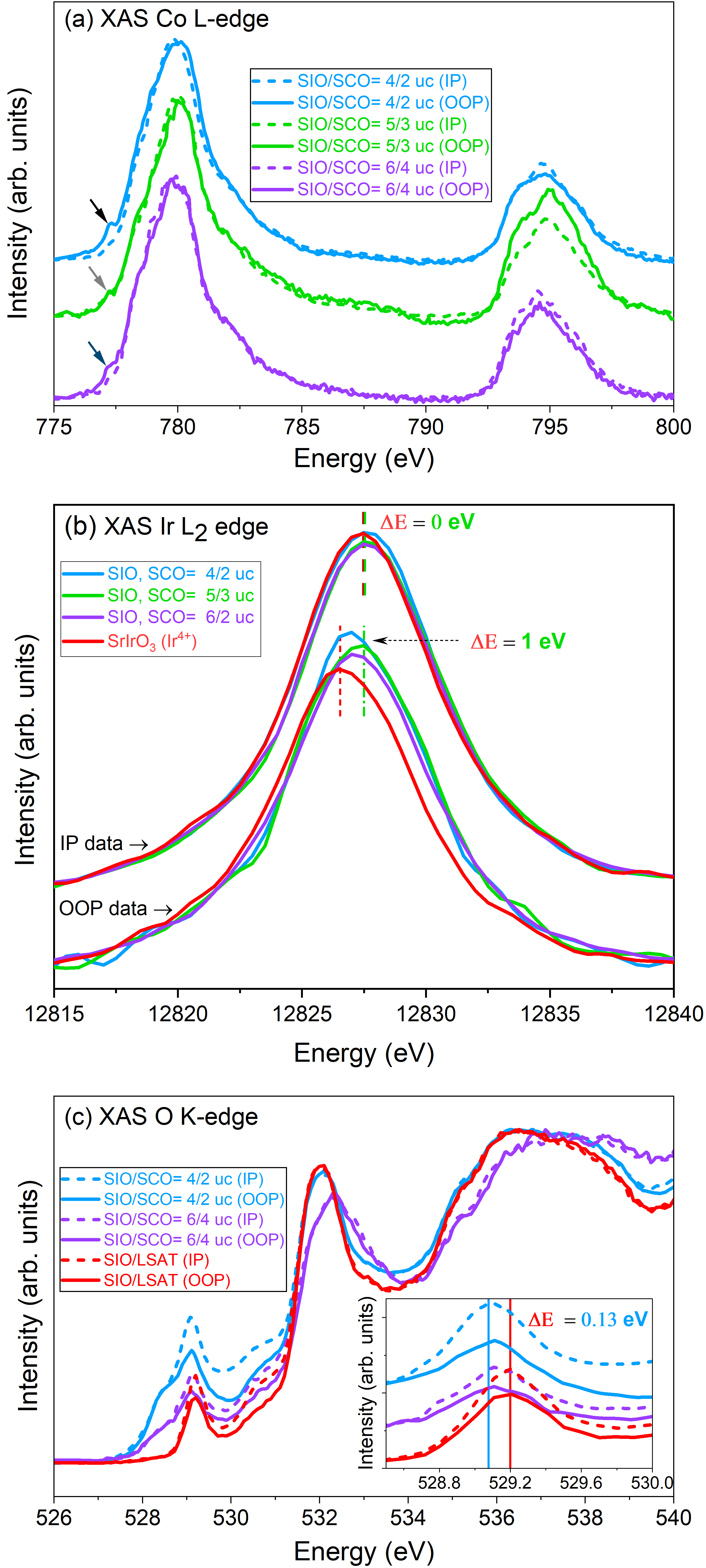}
\caption{Polarization dependent-XAS data of superlattices compared with a reference SrIrO$_3$ film: (a) Co L-edge, (b) Ir L$_2$-edge, and (c) O K-edge. All spectra were collected in the fluorescence yield detection mode.}
\label{XAS-Ir-L-Co-L-O-K}
\end{figure}
\clearpage

All out-of-plane (OOP) Co L-edge spectra exhibit features characteristic of Co$^{2+}$ as indicated by the arrows in Figure \ref{XAS-Ir-L-Co-L-O-K}(b)\cite{merz2010x,liao2019valence}. According to the dipole selection rules \(\Delta\)J = 0, ±1), the Co $L_2$-edge primarily probes transitions from 2p\textsubscript{1/2 }$\rightarrow$ 3d\textsubscript{3/2}, which correspond to the e\textsubscript{g} orbitals. In contrast, the Co $L_3$-edge includes transitions to both 3d\textsubscript{3/2} and 3d\textsubscript{5/2} states \cite{van1997anisotropic, george2024x}. With linearly polarized X-rays, the $L_2$-edge selectively probes different e\textsubscript{g} orbitals: in-plane (IP) polarization is sensitive to d$_{x^2-y^2}$ orbitals and out-of-plane (OOP) polarization probes d$_{3z^2-r^2}$orbitals \cite{esser2018strain}. Using the definition I($L_3$)/ [I($L_3$) + I($L_2$)], we obtained Branching ratio values for the superlattice samples with (4/2) uc, (5/3) uc and (6/4) uc using IP (OOP) XAS as follows: 0.704 (0.732), 0.693 (0.644), 0.695 (0.731) respectively. The polarization dependence of the branching ratio indicates a deviation from cubic symmetry \cite{van1997anisotropic}. This deviation is influenced by both the valence band spin-orbit coupling ⟨\textbf{L·S}⟩ and crystal field splitting, which are sensitive to Co-O-Co bond angles and Co-O bond lengths, modified by strain or doping \cite{merz2010x}. Such distortions would be expected due to both strain-induced octahedral distortions and charge-transfer-induced polar distortions in the system. Both structural distortions were predicted in our DFT models. 

In the case of Ir valence, the Ir  $L_2$-edge energies can be used to determine the oxidation state of Ir (Figure \ref{XAS-Ir-L-Co-L-O-K}a)\cite{laguna2015electronic, agrestini2018probing,sweers2024epitaxial, duell2021iridium}. OOP Ir $L_2$ data of the superlattice samples exhibit peak broadening that is found only in Ir$^{5+}$ and Ir$^{6+}$ states \cite{laguna2015electronic}. The Ir $L_2$ white-line shift associated with an oxidation state change from Ir$^{4+}$ from Ir$^{5+}$ is $\sim$1.2-1.3 eV \cite{laguna2015electronic, agrestini2018probing, duell2021iridium} suggesting an average state of close to Ir$^{5+}$ in the superlattice films based on OOP orbitals. We observe no noticeable shift between the SIO reference and the superlattice samples along the IP direction. We rule out strain as the origin of the peak shifts since the SIO reference is also grown on LSAT. 


X-ray linear dichroism (XLD) provides valuable insights into anisotropy of the electronic structure in superlattice films, where interfaces add an additional degree of symmetry-breaking beyond what may already be present from epitaxial strain \cite{disa2015orbital,comes2016interface}. Differences between in-plane and out-of-plane polarization may indicate orbital polarization \cite{disa2015orbital} due to charge transfer or polar distortions due to interfacial effects \cite{comes2016interface}. XLD for pure SIO in Figure \ref{sup-XLD-ALL-tog}(a) shows that IP Ir $L_2$ shifts towards higher energy by 1 eV relative to OOP Ir $L_2$, which is likely attributable to strain-induced effects due to compressive strain from the LSAT substrate. The transition intensity in the XAS $L$-edge is directly proportional to the number of empty states in the direction of \textbf{E}. Thus, the IP Ir $L_2$-edge probes $d_{x^2-y^2}$ and $d_{xy}$ orbitals. On the other hand, the $J_{eff}$ band in SIO undergoes a splitting due to a compressive strain and the resulting tetragonal distortion\cite{liu2013tuning}. Fig. \ref{sup-XLD-ALL-tog}(a) suggests that orbital splitting in SIO on both pure and superlattice structures is accompanied by a significant occupancy difference between in-plane and out-of-plane orbitals, referred to as orbital polarization \cite{benckiser2011orbital, disa2015orbital}. Given that charge transfer in the superlattice increases the number of available unoccupied J$_{eff}$ = 1/2 $t_{2g}$ states on Ir, we conclude that peak broadening in the OOP data indicates holes on Ir sit preferentially on the $d_{xz/yz}$ J$_{eff}$ = 1/2 orbitals.

In addition to the cation valence changes inferred from XAS, we also employed O $K$-edge XAS to examine changes in electronic band alignment due to charge transfer. The pre-edge feature in the O $K$-edge spectra arises from hybridization between O 2p and transition metal d orbitals- specifically, O 2p-Co 3d and O 2p-Ir 5d $t_{2g}$ \cite{pinta2008suppression,yang2020strain, chaurasia2021low, liu2017synthesis}. The intensity of this peak is proportional to the number of holes in the Co 3d and Ir 5d $t_{2g}$ states. The electron yield mode shows a more pronounced pre-edge peak due to its sensitivity to the near-surface region of the film. In contrast, bulk-sensitive fluorescence measurements include contributions from the substrate and show a less pronounced signal \cite{karvonen2010k}. However, pre-edge features for both samples were qualitatively similar and we obtained better signal to noise in fluorescence mode. Electron yield data are included in the Supplemental Information for comparison.

For the (SIO/SCO)=(4/2) uc and (6/4) uc superlattice films in Fig. \ref{XAS-Ir-L-Co-L-O-K}c, the pre-edge peak is centered around $\sim$529.1 eV. In contrast, the peak in a pure SIO film on LSAT appears at $\sim$ 529.2 eV, indicating a shift in the unoccupied hybridized O 2p states relative to the Fermi level. This is consistent with the DFT predictions, which showed that the O 2p band center within SIO would move $\sim$ 0.1-0.2 eV closer to the Fermi level. In both the SL and pure SIO films, the IP pre-edge peak is more pronounced than the OOP peak. This suggests that the combined hybridization involving (Ir d\textsubscript{xy} - planar O p\textsubscript{x,y}) and (Ir d\textsubscript{xz,yz} - apical p\textsubscript{x,y}) is stronger than the (Ir d\textsubscript{xz,yz} - planar O p\textsubscript{z}) orbitals \cite{liu2017synthesis,yang2020strain, chaurasia2021low}. The relative difference in intensity is greatest for the (SIO/SCO)=(4/2) superlattice where the interface density is highest, suggesting that interfaces amplify the anisotropy. A distinct shoulder peak at 528.4 eV is observed only in the superlattice films, and can be attributed to O 2p-Co 3d $t_{2g}$ hybridization \cite{pinta2008suppression,karvonen2010k,li2024emergence}. The position of the feature is consistent with a Co$^{3+}$ charge state, whereas Co$^{4+}$ would fall at lower photon energy ($\sim$527.5 eV)\cite{karvonen2010k}. Thus, we conclude that charge transfer from Ir to Co serves to hole-dope SIO, pinning valence band closer the Fermi level in those layers of superlattice. Meanwhile, Co is electron doped to a Co$^{3-\delta}$ charge state.




\section{Conclusions}
In conclusion, we have experimentally and theoretically demonstrated that Ir donates electrons to Co in the SrIrO$_3$/SrCoO$_3$ superlattices, thereby confirming the direction of charge transfer predicted for similar perovskite superlattices \cite{zhong2017band, chen2017charge}. Ir $L$-edge and O $K$-edge XAS data indicate orbital anisotropy in hybridized Ir 5d-O 2p states between in-plan and out-of-plan bonds due to the oxygen-mediated charge transfer at interfaces. We also demonstrate tuning of the O 2p band center in SIO layers via hole-doping, as confirmed by enhanced metallicity in SIO-SCO superlattices compared to uniform SIO and shifting of the O pre-edge peak position in the superlattices. Our findings not only offer a new perspective towards understanding and improving existing theories on charge transfer mechanism in metallic transition-metal oxides, but also demonstrate a viable strategy for tuning the properties of SIO and stabilizing P-SCO. Future works will explore the impacts of this charge transfer on magnetic behavior of these materials to understand the potential for emergent magnetic states that combine ferromagnetism and strong spin-orbit coupling.

\section{Data Availability}
Theory, spectroscopy, and diffraction data that support the findings of this article are openly available in text-based format through Zenodo at \href{http://doi.org/10.5281/zenodo.17517290}{DOI: 10.5281/zenodo.17517290}. Additional processed data is available from the corresponding author upon request.

\section{Acknowledgements}
This work was supported by the U.S. Department of Energy, Office of Science under contract number DE-SC0023478. The XRD was acquired through Major Research Instrumentation program from the National Science Foundation (NSF) under DMR-2018794. This research used resources in the Beamline for Materials Measurement (6-BM) of the National Synchrotron Light Source II, a U.S. Department of Energy (DOE) Office of Science User Facility operated for the DOE Office of Science by Brookhaven National Laboratory under Contract No. DE-SC0012704. We thank Bruce Ravel for training and assisting in running XAS measurements in the 6-BM. Use of the Advanced Photon Source, an Office of Science user facility, was supported by the U.S. Department of Energy, Office of Science, Office of Basic Energy Sciences under Contract No. DE-AC02-06CH11357. This work used Stampede3 at Texas Advanced Computing Center through NSF allocation DMR-110093 from the Advanced Cyberinfrastructure Coordination Ecosystem: Services \& Support (ACCESS) program, which is supported by NSF grants \#2138259, \#2138286, \#2138307, \#2137603, and \#2138296.

\clearpage

\renewcommand{\thefigure}{S\arabic{figure}}  
\renewcommand{\thetable}{S\arabic{table}}    
\renewcommand{\theequation}{S\arabic{equation}}  
\setcounter{figure}{0} 
\setcounter{table}{0}
\setcounter{section}{0}
\graphicspath{ {./Figures-sup/} }
\centering\textbf{Supplementary Information}

\section{Supplementary Crystal Structure of the Superlattice films}

\begin{figure}[htbp]
\centering
\includegraphics[width=1\textwidth]{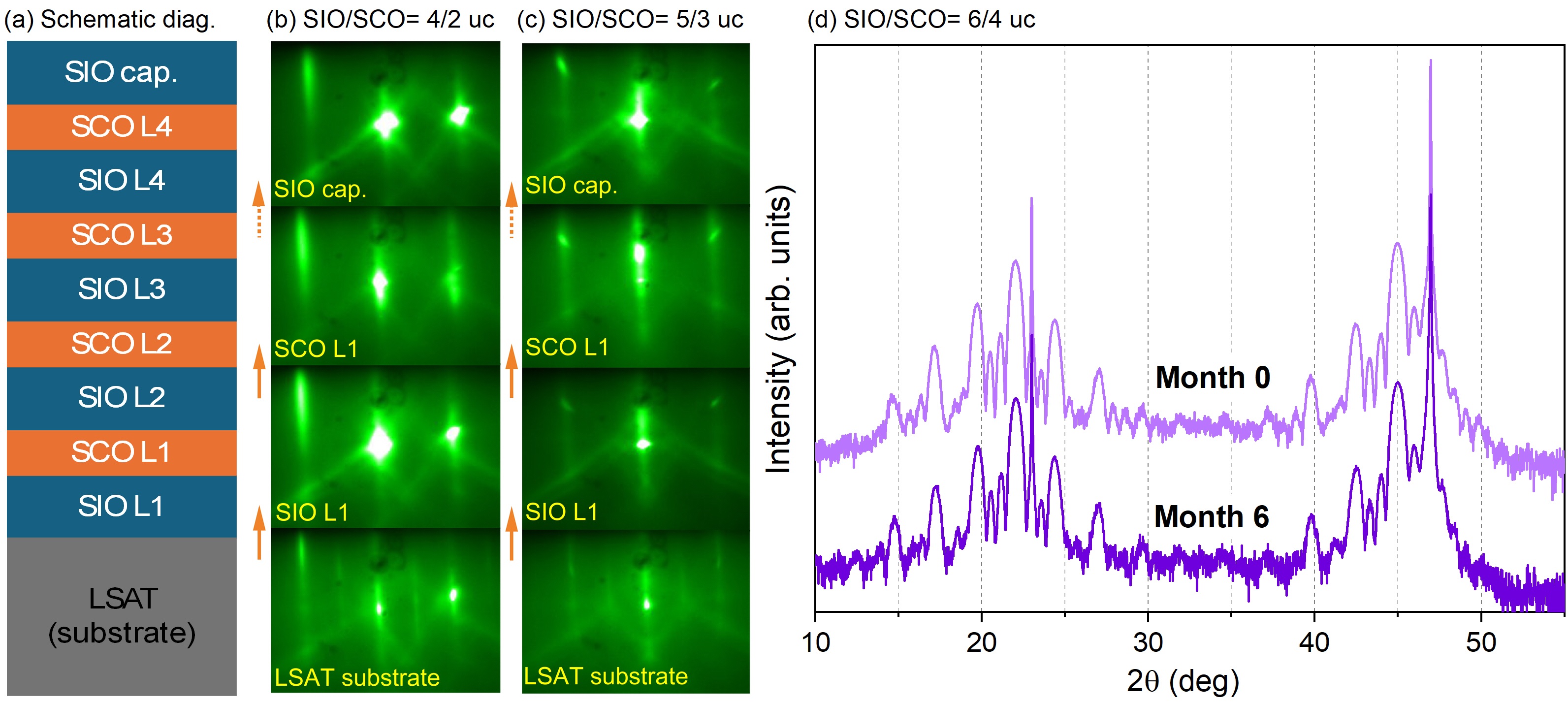}
\caption{(a) Schematic diagram of the superlattice (SL) films with each bilayer consisting of \textit{(SIO, SCO)= (4, 2)} unit cells (b= 6 uc) and \textit{(SIO, SCO)= (5, 3)} unit cells (b= 8 uc) (b)-(c) RHEED pattern of the corresponding SLs, (d) XRD evolution of SL (b= 10 uc) confirms that the structure remains stable over 6 months}
\label{sup-RHEED-SL1-SL2-and-XRD-SL3}
\end{figure}

\clearpage


\section{Density Functional Theory}
\begin{table}[htbp]
    \centering
\caption{Layer resolved orbital occupancy 20-atom heterostructures; Bold = Co layer}
\label{tab:DFT-orbital-20atm}
    \begin{tabular}{|l|c|c|c|}
        \hline     
        & Sr$_4$Ir$_2$O$_{12}$& Sr$_4$Co$_2$Ir$_2$O$_{12}$& Sr$_4$Co$_4$O$_{12}$\\
        & (5x5x4; IS=-5) & (5x5x4; IS=-5) & (5x5x4; IS=-5) \\
        \hline
        $x_{Co}$   & 0.00 & 0.50 & 1.00 \\
        \hline
        0.750, OOP     & 0.545 & 0.567 & 0.632 \\
        \hline
        0.500, IP& 6.562; 0.535 & \textbf{6.785; 0.496} & \textbf{7.154; 0.619} \\
        \hline
        0.250, OOP     & 0.545 & 0.567 & 0.632 \\
        \hline
        0.000, IP      & 6.562; 0.535 & 6.245; 0.496 & \textbf{7.154; 0.619} \\
        \hline
    \end{tabular}
\end{table}


\begin{table}[htbp]
    \centering
\caption{Layer resolved orbital occupancy, 40-atom heterostructures; Bold = Co layer}
\label{tab:DFT-orbital-40atm}
    \begin{tabular}{|l|r|r|r|r|r|}
        \hline
        & Sr$_8$Ir$_8$O$_{24}$ & Sr$_8$Co$_2$Ir$_6$O$_{24}$ & Sr$_8$Co$_4$Ir$_4$O$_{24}$ & Sr$_8$Co$_6$Ir$_2$O$_{24}$ & Sr$_8$Co$_8$O$_{24}$ \\
        & & & & & \\
        \hline
 $x_{\text{Co}}$& 0.00& 0.25& 0.50& 0.75&1.00\\
        \hline
        0.875, OOP & 0.557 & 0.575 & 0.570 & 0.604 & 0.595 \\
        \hline
        0.750, IP & 6.751; 0.591 & 6.669; 0.586 & 6.483; 0.647 & \textbf{6.864; 0.585} & \textbf{6.838; 0.591} \\
        \hline
        0.625, OOP & 0.557 & 0.575 & 0.534 & 0.542 & 0.595 \\
        \hline
        0.500, IP & 6.751; 0.591 & 6.589; 0.608 & \textbf{6.892; 0.554} & \textbf{6.824; 0.525} & \textbf{6.838; 0.591} \\
        \hline
        0.375, OOP & 0.557 & 0.529 & 0.534 & 0.520 & 0.595 \\
        \hline
        0.250, IP & 6.751; 0.591 & \textbf{7.002; 0.446} & \textbf{6.892; 0.554} & \textbf{6.864; 0.585} & \textbf{6.838; 0.591} \\
        \hline
        0.125, OOP & 0.557 & 0.635 & 0.403 & 0.488 & 0.595 \\
        \hline
        0.000, IP & 6.751; 0.591 & 6.589; 0.621 & 6.483; 0.647 & 6.294; 0.673 & \textbf{6.838; 0.591} \\
        \hline
    \end{tabular}

\end{table}

\section{Sr$_2$CoIrO$_{6}$ Growth and Characterization}
\raggedright
We grew Sr$_2$CoIrO$_{6}$ (SCIO) films on SrTiO$_{3}$ (STO) (100) substrate obtained from MTI corporation. We sonicated each STO substrate (as well as LSAT substrate for superlattice films) in Acetone and Isopropyl alcohol for 10 minutes. We found a relatively wide window of growth temperature for the double perovskites (625 – 700 °C), whereas the superlattices had in a much narrower range (615–635 °C).  We ensured a 2:1 ratio between Sr and Co using Rutherford Backscattering Spectrometry (RBS). The STO substrates were annealed in oxygen plasma for 30 minutes at 800 °C. SCIO films were grown at 625 °C and all three sources were supplied simultaneously. An oxygen flow rate of 1.0 sccm (P= $6 \times 10^{-5}$ Torr) was maintained during growth. It’s worth noting that, this relatively high oxygen pressure was essential for growing SCIO films, however it contributed to the oxidation of the Ir source material and leading to it’s rapid depletion. Because of this we grew only relatively a small number of SCIO films

\begin{figure}[t]
\centering
\includegraphics[width=\textwidth]{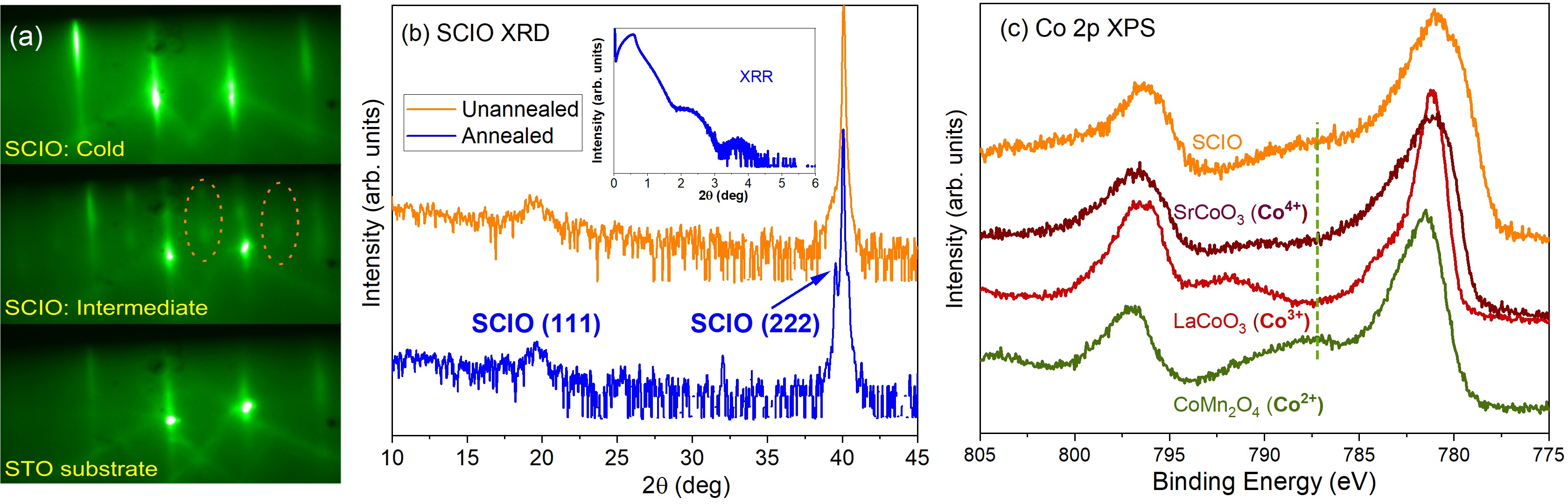}
\caption{(a)-(b) RHEED and off-normal XRD of SCIO films showing cation ordering, (c) Co 2p XPS of Sr\textsubscript{2}CoIrO\textsubscript{6} compared with cobaltate reference compounds grown in the same MBE.\cite{blanchet2021electronic}}
\label{sup-SCIO-structure}
\end{figure}

To promote cation ordering and investigate any change in electronic structure, one of the SCIO films was post-annealed at 750 °C for 3 hours in a flowing air tube furnace. Because of a small difference in cation radii (0.04 A) between Co$^{3+}$ and Ir$^{5+}$\cite{esser2018strain}, achieving a spontaneously ordered phase in SCIO was inherently challenging \cite{ohtomo2013spontaneous,anderson1993b}. Although we didn’t quantitatively determine the ordering parameter, off-axis XRD shown in Figure \ref{sup-SCIO-structure}(b) reveals the presence of superlattice reflections at half-integer Bragg peaks consistent with the double perovskite SCIO (111) peak. This indicates the formation of an ordered superstructure, providing evidence of spontaneous B-site ordering in our double perovskites\cite{esser2018strain,manako1999epitaxial}. Furthermore, in-situ RHEED exhibits two-fold superstructure streaks (highlighted by the dashed orange ellipse) along azimuthal (110) that corresponds to the double perovskite unit cell\cite{manako1999epitaxial, hashisaka2006epitaxial} 
\clearpage

\begin{figure}[htbp]
\centering
\includegraphics[width=\textwidth]{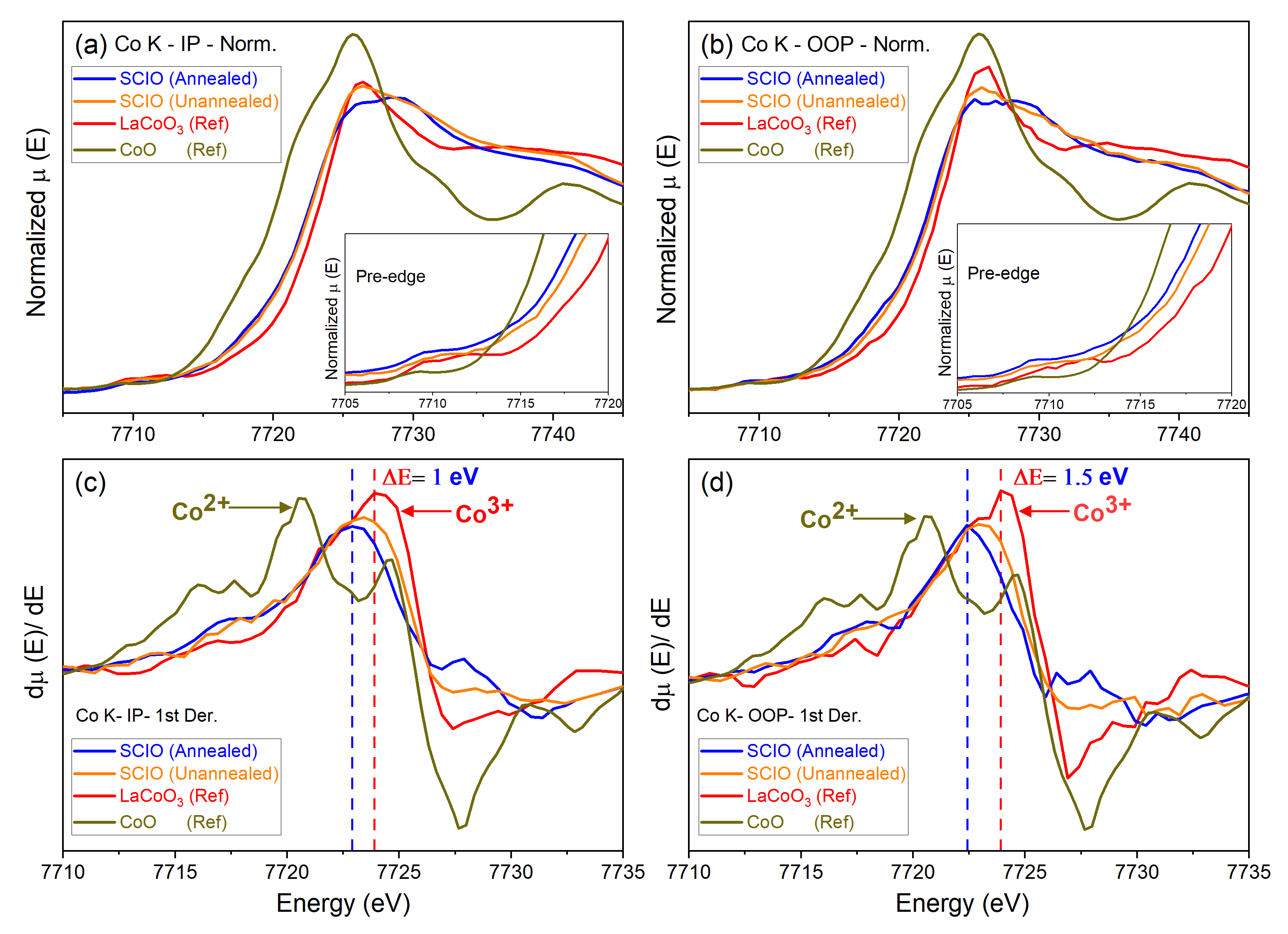}
\caption{XAS Co K-Edge data for the double derovskite films (a), (c) In-plane data (normalized spectra and their first derivatives), (b)(d) Out-of-plane data (normalized spectra and their first derivatives).}
\label{sup-XAS_SCIO}
\end{figure}

\begin{figure}[htbp]
\centering
\includegraphics[width=0.5\textwidth]{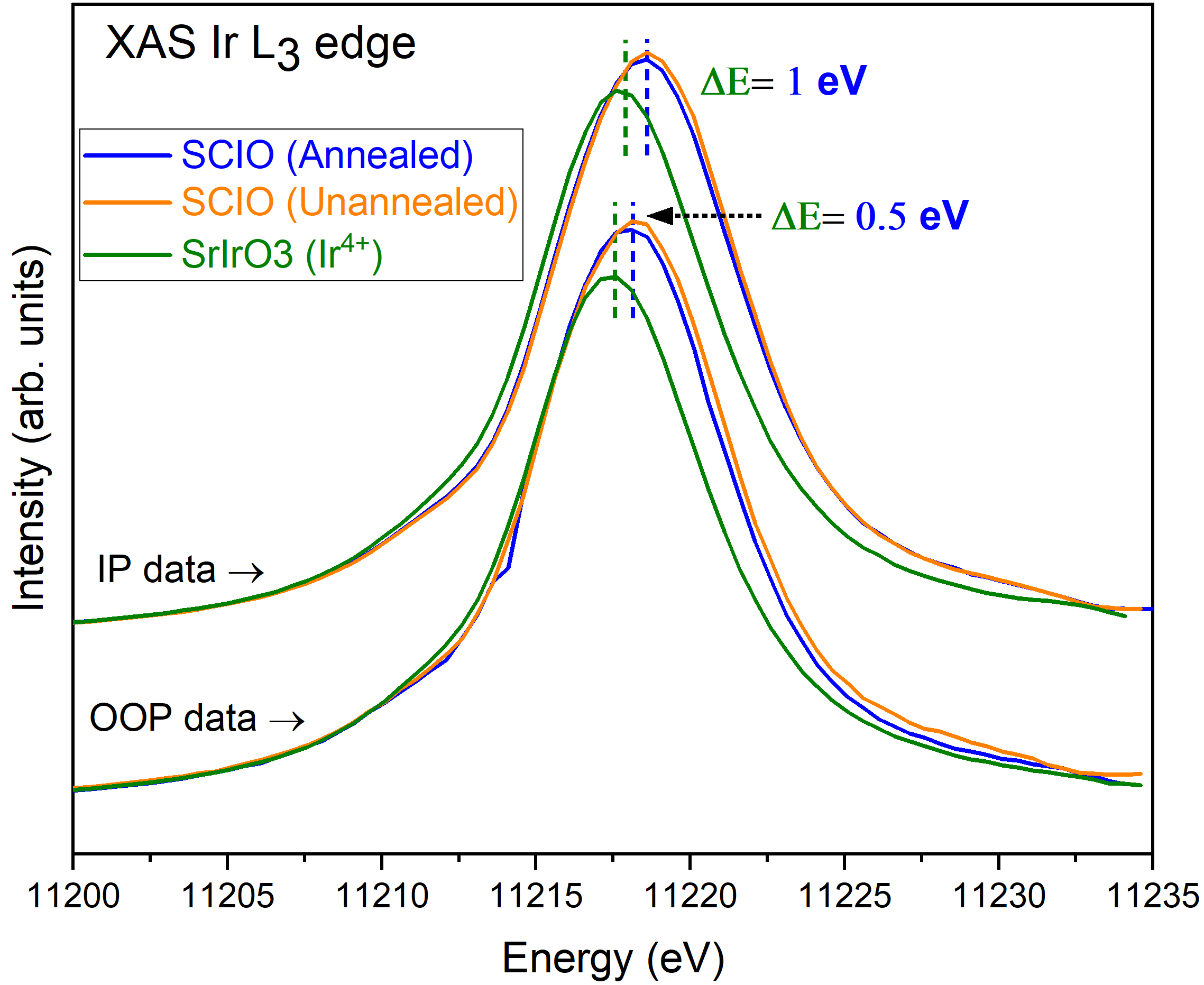}
\caption{Polarization-dependent XAS at Ir L\textsubscript{3} edge of Sr\textsubscript{2}CoIrO\textsubscript{6} samples with a reference SrIrO\textsubscript{3} film. All films were grown on SrTiO3 substrate}
\label{sup-XAS Ir-L3}
\end{figure}

\clearpage
\section{X-ray linear dichroism (XLD)}

\begin{figure}[h]
\centering
\includegraphics[width=1\textwidth]{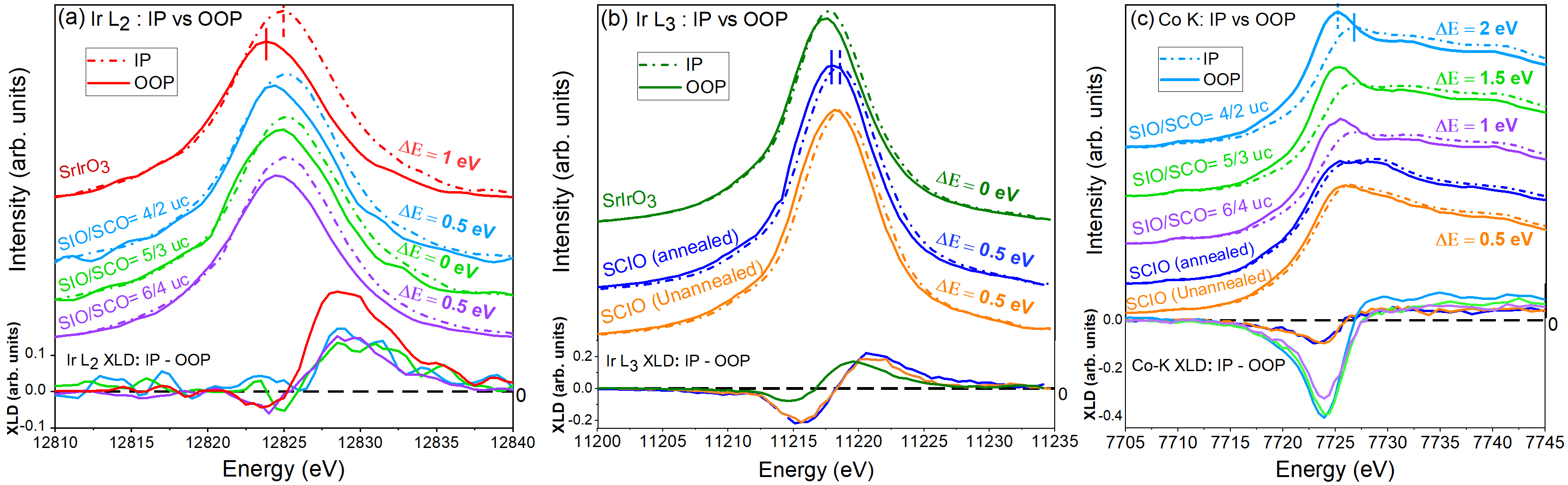}
\caption{X-ray linear dichroism data at (a) Ir L\textsubscript{2} edge of superlattices, (b) Ir L\textsubscript{3} edge of double perovskites (c) Co K edge of the superlattices.}
\label{sup-XLD-ALL-tog}
\end{figure}


\section{Co K Pre-edge fitting} 
\raggedright

\begin{figure}[htbp]
\centering
\includegraphics[width=1\textwidth]{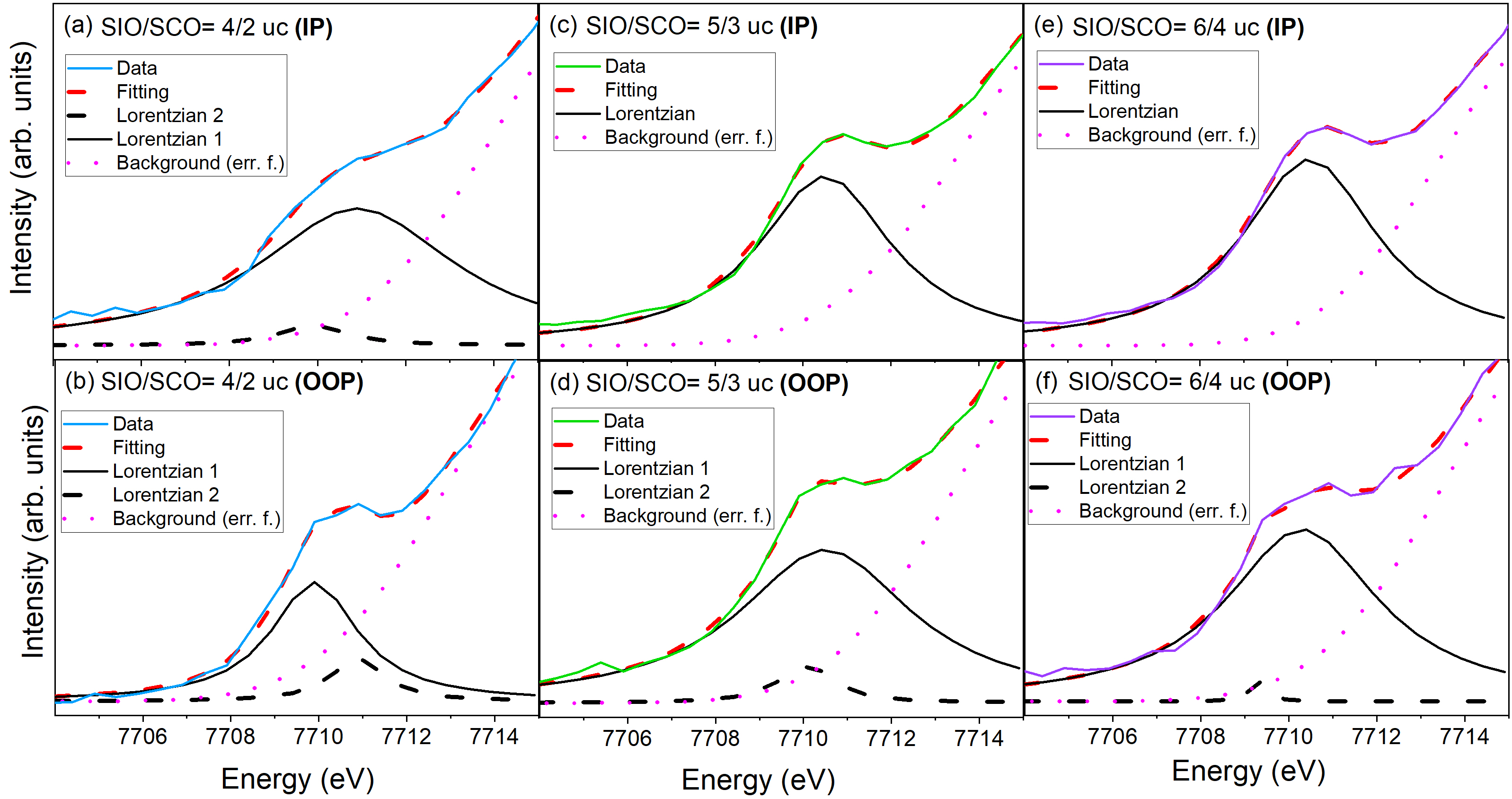}
\caption{Fitting of the Co K Pre-edge data of the superlattice films (performed in the Athena software).}
\label{sup-pre-edge-fitting-SL}
\end{figure}

\begin{figure}[htbp]
\centering
\includegraphics[width=1\textwidth]{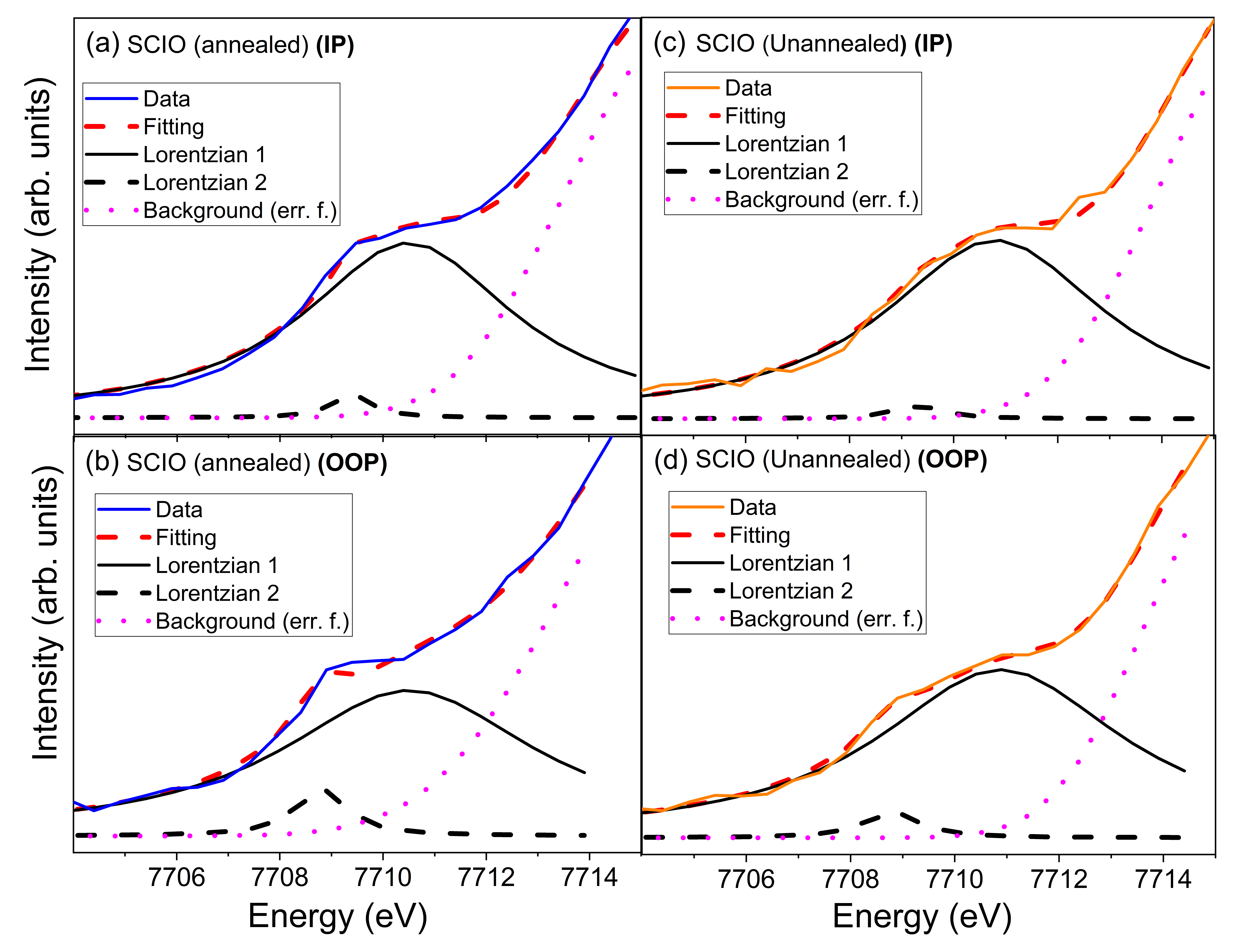}
\caption{Fitting of the Co K Pre-edge data of the SCIO films (performed in the Athena software).}
\label{sup-Pre-edge-fitting-SCIO}
\end{figure}

We analyzed polarization-dependent pre-edge data to investigate potential 4p-3d mixing in Co and octahedral distortions in SCO layers of the superlattices. In a perfect octahedral system, 4p-3d mixing is forbidden because of the inversion symmetry, and any pre-edge feature arises solely due to quadrupolar contribution \cite{george2024x, de20091s}.  Any pre-edge feature is indicative of dipolar contributions, which indicates 4p-3d orbital mixing. Therefore, the degree of 4p–3d mixing can be determined from the dipolar contribution, which is expected to follow a \(\cos^2 \theta \) angular dependence \cite{yano2007polarized, rossi2019x, brouder1990angular, joseph2010study}. To analyze the polarization dependence of the pre-edge features, we fitted the peaks using Lorentzian functions and modeled the background with an error function as shown in Figs. \ref{sup-pre-edge-fitting-SL}-\ref{sup-Pre-edge-fitting-SCIO} and Table \ref{pre-edge fitting table}. For Co, the quadrupolar contribution accounts for only about 4\% of the dipolar component in the pre-edge intensity\cite{george2024x}, consequently even a moderate dipolar contribution should appear in a significantly enhanced manner.

\begin{table}[htbp]
\centering
\caption{Co K Pre-edge fitting results.}
\label{pre-edge fitting table}
\begin{tabular}{|l|ccc|ccc|}
\hline
& \multicolumn{3}{c|}{Primary Peak} & \multicolumn{3}{c|}{Sec. Peak} \\
\hline
& Center (eV) & Intensity & Sigma & Center (eV) & Intensity & Sigma \\
\hline
SIO/SCO= 4/2 uc \textbf{(IP)}& 7710.90 & 0.41  & 5.44 & 7709.90 & 0.02  & 1.78 \\ \hline
SIO/SCO= 4/2 uc \textbf{(OOP)}& 7709.90 & 0.156 & 2.37 & 7710.90 & 0.035 & 1.44 \\ \hline
SIO/SCO= 5/3 uc \textbf{(IP)}& 7710.50 & 0.343 & 3.67 & 7709.40 & 0.033 & 1.47 \\ \hline
SIO/SCO= 5/3 uc \textbf{(OOP)}& 7710.05 & 0.394 & 4.37 & 7710.05 & 0.031 & 1.43 \\ \hline
SIO/SCO= 6/4 uc \textbf{(IP)}& 7710.05 & 0.38  & 3.69 & 7709.05 & 0.02  & 1.46 \\ \hline
SIO/SCO= 6/4 uc \textbf{(OOP)}& 7710.30 & 0.396 & 4.77 & 7709.30 & 0.005 & 0.329 \\ \hline
SCIO (Annealed) \textbf{(IP)} & 7710.50 & 0.41  & 4.97 & 7709.40 & 0.01  & 0.849 \\ \hline
SCIO (Annealed) \textbf{(OOP)} & 7710.05 & 0.355 & 4.55 & 7708.95 & 0.01  & 0.783 \\ \hline
SCIO (Unannealed) \textbf{(IP)} & 7709.90 & 0.435 & 5.18 & 7709.25 & 0.007 & 1.10 \\ \hline
SCIO (Unannealed) \textbf{(OOP)} & 7710.90 & 0.451 & 5.72 & 7708.80 & 0.015 & 1.22 \\
\hline
\end{tabular}
\end{table}

\begin{figure}
    \centering
    \includegraphics[width=1\linewidth]{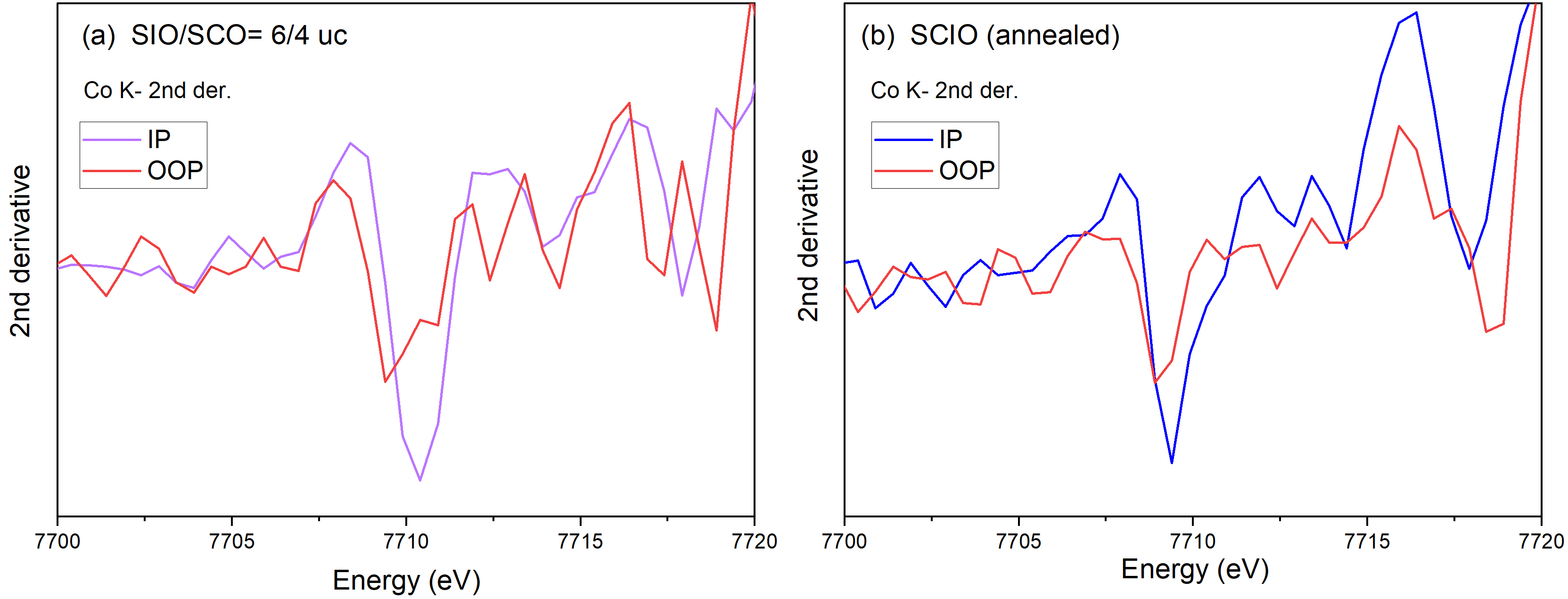}
    \caption{Second derivative of the Co K-pre edge spectra of representative SL and SCIO films}
    \label{sup-pre-edge-2nd-der}
\end{figure}
\clearpage

\section{XAS Co L-edge and O K-edge}
\raggedright

\textbf{Co L-edge and O K-edge data processing}: A linear pre-edge background subtraction was performed in Both Co L and O K-edge data. The Co L-edge data was further treated with a background subtraction using an arctangent function. Following background removal, each Co L and O K-edge data was normalized by their respective maximum intensity. To determine Branching ratio of the L edge: L3 is taken between 774-785 eV, L2 is taken between 791-800 eV

\begin{figure}[htbp]
   \centering
    \includegraphics[width=1\linewidth]{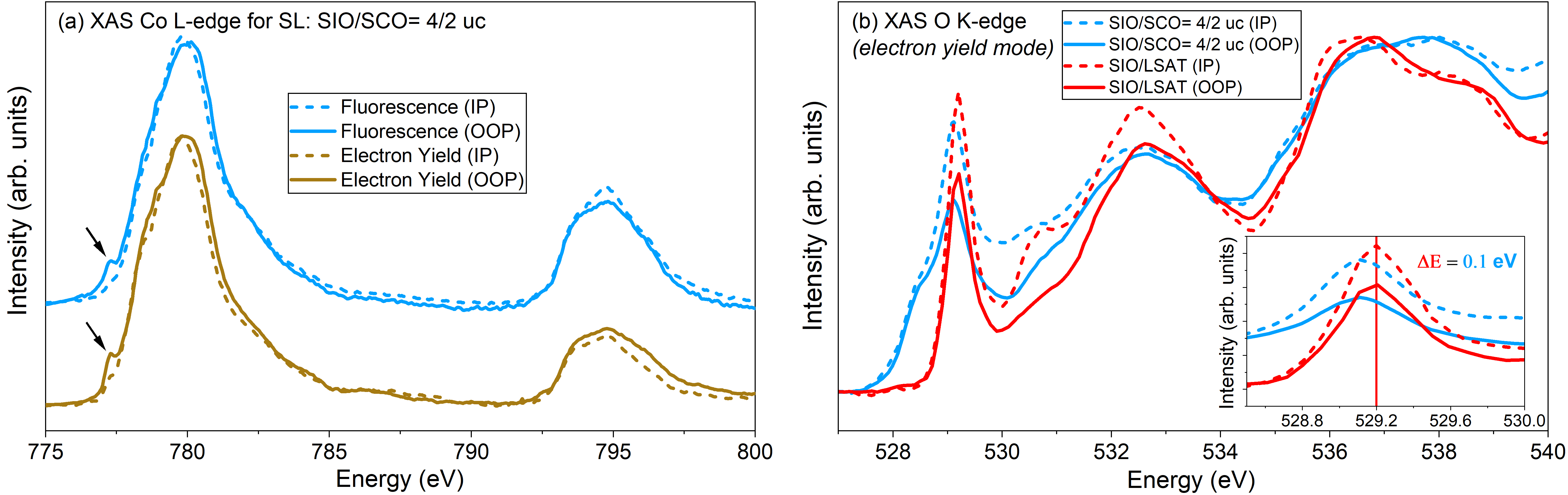}
    \caption{Polarization-dependent XAS at (a) Co L and (a) O K-edges for the superlattice film SIO/SCO= 4/2 uc}
    \label{sup-XAS-Co-L-O-K-EY}
\end{figure}

\section{Electrical Transport Methods}
\raggedright
The electrical transport property measurements of superlattices were performed in a Quantum Design Physical Property Measurement System (PPMS) DynaCool. We used the DC four-probe method in a van der Pauw geometry with indium contacts placed at the four corners of the films to serve as Ohmic contacts. For temperature-dependent studies, the system was allowed to stabilize for one hour after reaching each target temperature to avoid any temperature lag and ensure accurate readings. 
\clearpage
\bibliography{Main-refs}

\end{document}